\documentclass[aps,pra,preprint,showpacs,superscriptaddress]{revtex4}
\usepackage{mathtools}
\usepackage{graphicx}
\usepackage{amssymb}
\usepackage{multirow}

\begin{document}
\title{\textsc{xcalib}: a focal spot calibrator for intense X-ray free-electron laser pulses 
based on the charge state distributions of light atoms}
\author{Koudai Toyota}
\affiliation{Center for Free-Electron Laser Science, DESY, 22607 Hamburg, Germany}

\author{Zoltan Jurek}
\affiliation{Center for Free-Electron Laser Science, DESY, 22607 Hamburg, Germany}

\author{Sang-Kil Son}
\affiliation{Center for Free-Electron Laser Science, DESY, 22607 Hamburg, Germany}

\author{Hironobu Fukuzawa}
\affiliation{Institute of Multidisciplinary Research for Advanced Materials, Tohoku University, Sendai, Japan}

\author{Kiyoshi Ueda}
\affiliation{Institute of Multidisciplinary Research for Advanced Materials, Tohoku University, Sendai, Japan}

\author{Nora Berrah}
\affiliation{Physics Department, University of Connecticut, Storrs, CT, USA}

\author{Benedikt Rudek}
\affiliation{Physikalisch-Technische Bundesanstalt, Braunschweig, Germany}

\author{Daniel Rolles}
\affiliation{J. R. Macdonald Laboratory, Department of Physics, Kansas State University, Manhattan, KS, USA}

\author{Artem Rudenko}
\affiliation{J. R. Macdonald Laboratory, Department of Physics, Kansas State University, Manhattan, KS, USA}

\author{Robin Santra}
\email{robin.santra@cfel.de}
\affiliation{Center for Free-Electron Laser Science, DESY, 22607 Hamburg, Germany}
\affiliation{Department of Physics, University of Hamburg, 20355 Hamburg, Germany}

\date{\today}

\begin{abstract}
We develop the \textsc{xcalib} toolkit to calibrate the beam profile of 
an X-ray free-electron laser (XFEL) at the focal spot based on
the experimental charge state distributions (CSDs) of light atoms.
Accurate characterization of the fluence distribution at the focal spot 
is essential to perform the volume integrations of physical
quantities for a quantitative comparison between theoretical and experimental results,
especially for fluence dependent quantities. The use of the CSDs of light atoms is 
advantageous because CSDs directly reflect experimental conditions at the focal spot,
and the properties of light atoms have been well established in both theory and 
experiment. To obtain theoretical CSDs, we use \textsc{xatom}, 
a toolkit to calculate atomic electronic structure and to simulate
ionization dynamics of atoms exposed to intense XFEL pulses,
which involves highly excited multiple core hole states.
Employing a simple function 
with a few parameters, the spatial profile of an XFEL beam 
is determined by minimizing 
the difference between theoretical and experimental results. We have implemented 
an optimization procedure employing the reinforcement learning technique.
The technique can automatize and organize calibration procedures which, before, had been performed manually.
\textsc{xcalib} has high flexibility, simultaneously 
combining different optimization methods,  
sets of charge states, and a wide range of parameter space.
Hence, in combination with \textsc{xatom}, 
\textsc{xcalib} serves as a comprehensive tool to calibrate 
the fluence profile of a tightly focused XFEL beam in the interaction region. 

\end{abstract}
\maketitle
\section{Introduction}
The recent X-ray free-electron laser (XFEL) technologies have enabled us to 
conduct experiments at ultrashort time scales ($\approx$ a few~fs) and ultrahigh 
intensities ($\approx 10^{20}$~W/cm$^2$), which are far beyond 
the domain of conventional synchrotron radiation sources \cite{Schneider2010}. 
Theoretical calculations have played a crucial role in revealing new ionization 
mechanisms of atoms and molecules found in experiments driven by such unprecedented 
light \cite{young2010,doumy2011,rudek2012,rudek2013,fukuzawa2013,motomura2013,murphy2014,rudenko2017,rudek2018}. In these discoveries, 
a quantitative comparison between theoretical and experimental results was
crucial to elucidate the underlying physics.
When an XFEL beam is focused onto a target in experiments,  
the fluence values of the beam have a non-uniform spatial 
distribution \cite{barty2009,schneider2016,nagler2017,schneider2018} in the focal spot,
so that a range of fluence values covered 
by the distribution may contribute to the yield of an observable such as ions, electrons, and photons. Therefore, 
when theoretically computing the yield of an observable, we need to add up 
all fluence-dependent contributions in order to make a comparison with experimental data, 
which is called volume integration~\cite{young2010}.
It is thus essential to calibrate the spatial fluence distribution in the 
focal spot to perform the volume integration. 
However, a direct measurement of the focal volume parameters for XFELs represents a significant experimental challenge (see, e.g., \cite{Chalupsky11}).

In previous studies \cite{young2010,doumy2011,rudek2012,rudek2013,fukuzawa2013,motomura2013,murphy2014,rudenko2017}, 
spatial fluence distributions were calibrated utilizing experimental and theoretical charge state 
distributions (CSDs) of light atoms such as neon (Ne) or argon (Ar) atoms. 
Recently, a calibration procedure at low and intermediate fluences based on fragment ion spectra of Ar clusters has been proposed in \cite{Kumagai18}.
The CSDs of Ne and Ar atoms are often used as fast experimental feedback for minimizing the focal spot size when changing the focusing mirror settings~\cite{Schorb12}.
This approach using CSDs of light atoms has three advantages. 
First, because of the high non-linearity of the XFEL interactions with atoms, these CSDs are very sensitive to the peak fluence value as well as to the spatial fluence profile in the focal spot.
Second, we can utilize the well established atomic properties of light atoms. 
Third, calculating the CSDs of light atoms is 
computationally cheap. The CSDs of atoms were calculated using the \textsc{xatom}
toolkit \cite{xatom2018}.
Assuming a specific functional form of the spatial fluence 
profile at the focal spot depending on a few parameters, the volume-integrated 
theoretical CSD is calculated. So far, the parameters have been determined by minimizing a 
certain measure by manually exploring the parameter space. However, such manual 
procedures lack efficient algorithms to obtain an optimized solution, and are 
insufficient to handle a large number of experimental results in a wide range of parameter space.
The situation motivated us to develop a toolkit to automatize the optimization procedures, 
employing the reinforcement learning technique \cite{raschka2015}.
A machine learning technique has been used in single-shot characterization of spectral 
and temporal profiles of XFEL pulses~\cite{sanchez-gonzalez2017}.
These optimization methods have an advantage in finding the direction
to a solution in the parameter space with efficient algorithms. 
We designed the \textsc{xcalib} toolkit to have flexibility simultaneously 
combining different pulse profiles, parameter ranges, 
charge states and optimization methods. Therefore, \textsc{xcalib} offers a comprehensive 
tool to calibrate X-ray beam parameters in XFEL experiments. 

The paper is organized as follows. 
In Sec.~\ref{sec:num}, we introduce our numerical method focusing on
the volume integration and the optimization method. In Sec.~\ref{sec:res}, we revisit
the three Ar calibrations in \cite{fukuzawa2013,murphy2014,rudek2018} 
to show that calibrated results by \textsc{xcalib} are consistent with the previous results. 
We also study the effect of attenuators on the fluence profile used in \cite{rudek2018}.
In Sec.~\ref{sec:conc}, we conclude the paper with a summary.
We use atomic units throughout the paper unless stated otherwise.

\section{Numerical method}
\label{sec:num}
\subsection{Volume integration}
In this section, we first formulate the numerical procedure for the volume integration for ion yield distributions,
which is essential to obtain theoretical results that may be compared to experimental results.
The ion yield $\mathcal{Y}_{\rm theo}^{(+q)}$ of charge state $q$ produced at 
position ${\bf r}$ in the focal spot is assumed to be a function 
of the position-dependent fluence value $F({\bf r};{\bf P})$,
where a set of parameters ${\bf P}$ characterizes the spatial profile.
The volume-integrated yield is then given by the following three-dimensional integral,
\begin{equation}
Y^{(+q)}_{\rm theo}({\bf P})
=\int \mathcal{Y}_{\rm theo}^{(+q)}\Bigl(F({\bf r;{\bf P}})\Bigr)d^3r.
\label{eq:absyield}
\end{equation}
We normalize $Y^{(+q)}_{\rm theo}({\bf P})$ as follows:
\begin{equation}
y^{(+q)}_{\rm theo}({\bf P})=\frac{1}{N({\bf P})}Y^{(+q)}_{\rm theo}({\bf P}),
\label{eq:ionyield}
\end{equation}
where the constant $N({\bf P})$ is a normalization factor given by
\begin{equation}
N({\bf P})=\sum_{{\rm all}~q} Y^{(+q)}_{\rm theo}({\bf P}).
\label{eq:norm}
\end{equation}
The summation runs over the set of all available charge states.
Hence, the sum over the normalized ion yields is unity,
\begin{equation}
\sum_{{\rm all}~q} y^{(+q)}_{\rm theo}({\bf P})=1.
\label{eq:norm_cond}
\end{equation}
In the following, ion yields refer to the normalized ion yield defined in Eq.~(\ref{eq:ionyield}), unless specified otherwise.

The theoretical ion yields $\mathcal{Y}_\text{theo}^{(+q)}$ 
are calculated by employing the \textsc{xatom} toolkit. \textsc{xatom} is a set 
of computer codes to calculate electronic structure of atoms based on the 
Hartree-Fock-Slater (HFS) method and to describe multiphoton multiple 
ionization dynamics during intense XFEL pulses employing a rate-equation approach. 
\textsc{xatom} has been tested with a series of gas-phase atomic experiments and has played 
a crucial role in many XFEL applications (see \cite{jurek2016} and references therein). 
Note that the theoretical ion yields can also be obtained by other tools, 
for example, 
\textsc{scfly}~\cite{chung2007,Ciricosta11},
\textsc{averro{\`e}s/transpec}~\cite{Peyrusse00,Peyrusse14},
\textsc{dlayz}~\cite{xiang2012,Gao13,Gao14}, and
\textsc{mcre}~\cite{ho2014,Ho15}.

\subsection{Single Gaussian spatial profile}
\begin{figure}
\includegraphics[scale=0.7]{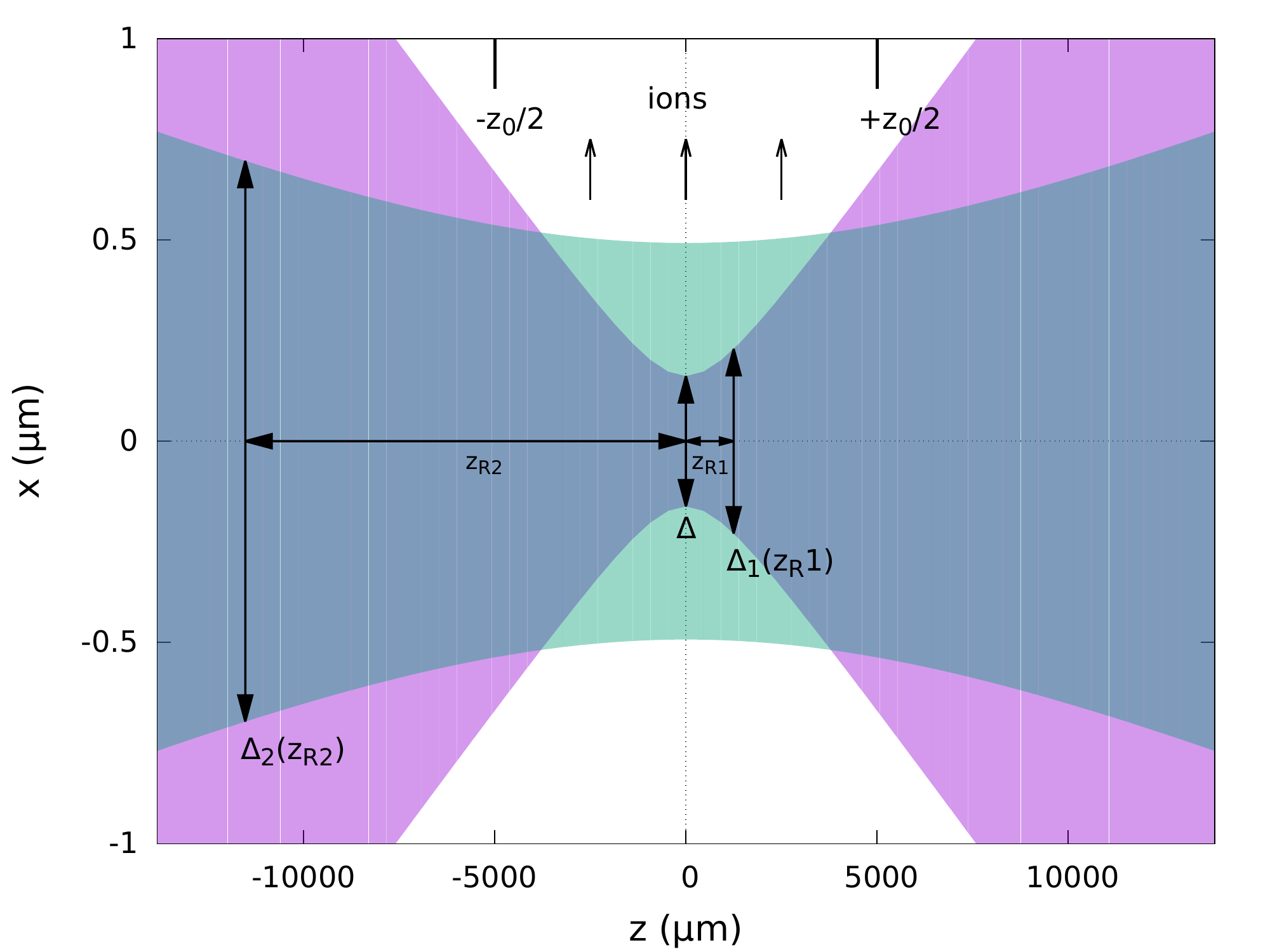}
\caption{\label{fig:beam_geo} 
The beam geometry in the $xz$ plane.
The X-ray beam direction is along the $z$ axis, and the target gas jet flows along the $x$ axis.
When the highest peak fluence is desirable, the target gas jet is located at $z=0$.
For the single Gaussian spatial profile, the beam shape in the $xz$ plane is shown in purple.
For the double Gaussian spatial profile, the purple area is for the first Gaussian and the green area for the second Gaussian.
$z_{R1}$ and $z_{R2}$ are the Rayleigh ranges for the first and second Gaussian profiles, respectively.
$\Delta$ indicates the X-ray beam width or focal spot size, which varies along with $z$.
The spatial fluence distribution is given in the $xy$ plane, and a spatial profile as a function of $x$ at $y=z=0$ is shown in Fig.~\ref{fig:beam_prof}.
The produced ions are collected by a detector with a slit length $z_0$ shown at the top of the figure.}
\end{figure}

To perform the volume integration, Eq.~(\ref{eq:absyield}), we need to model the 
spatial fluence profile, $F({\bf r};{\bf P})$, to map a given position ${\bf r}$ 
to a fluence value in the interaction volume. The interaction volume is defined 
by the intersection 
between the XFEL beam and the target gas jet (atomic or molecular beam). 
If the target beam size (typically $\sim$mm) is larger than the XFEL 
beam size ($\sim$$\mu$m or less), the XFEL beam determines the shape of the intersection in the direction transverse to the XFEL beam propagation.
Since the tightly focused XFEL beam diverges with increasing distance $z$ from the focus, the shape of the interaction 
volume is similar to an hourglass laid on the $z$-axis, 
as shown in purple in Fig.~\ref{fig:beam_geo}. 
Assuming Gaussian beam optics, the diameter of the cross 
section transverse to the $z$-axis is given by
\begin{equation}
\Delta_1(z) = \Delta \sqrt{ 1 + \left( \frac{z}{z_{R1}} \right)^2 }, 
\label{eq:bsize1}
\end{equation}
where the quantity $\Delta$ represents the focal spot size, and
$z_{R1}$ represents the Rayleigh range. 
Let $\lambda$ be the wavelength of the XFEL beam.
The Rayleigh range is then given by
\begin{equation}
z_{R1}=\frac{1}{2\ln(2)}\frac{\pi \Delta^2}{\lambda}.
\label{eq:zr1}
\end{equation} 
The diameter of the cross section at $z=z_{R1}$ is $\sqrt{2}$ 
times larger than that at $z=0$, namely, $\Delta_1(z=z_{R1})=\sqrt{2} \Delta$.
The beam geometry is depicted in Fig.~\ref{fig:beam_geo}.
The Rayleigh range becomes larger as the photon energy increases and smaller as the focal size decreases.
We model the spatial fluence distribution transverse to the XFEL beam propagation direction 
(the $z$-axis in Fig.~\ref{fig:beam_geo}).
In this work, we employ two types of spatial fluence profile. 
One of them is a single Gaussian spatial profile (SGSP) given by
\begin{equation}
F({\bf r};F_0)=\frac{\Delta^2}{\Delta^{2}_{1}(z)}
F_0e^{-\pi a\frac{x^2+y^2}{\Delta_{1}^2(z)}}.
\label{eq:sgsp3d}
\end{equation}
The constant $a=4{\rm ln}(2) / \pi$ is chosen so that
$\Delta_{1}(z)$ becomes the full-width at half-maximum (FWHM). 
The beam size $\Delta_{1}(z)$ is given by Eq.~(\ref{eq:bsize1}).
The quantity $F_0$ represents the peak fluence defined by 
number photons per unit area. Because it was found in \cite{barty2009} that
the spatial profile has a shape similar to a Gaussian function, 
we also model the spatial profile employing Gaussian functions in this work. 
The model allows us to reduce the number of calibration parameters, which reduces
the computational effort. 
The total number of photons $n$ in the $xy$-plane at an arbitrary value of $z$ 
is given by
\begin{equation}
n=\int F({\bf r};F_0)dxdy=\frac{F_0\Delta^2}{a}.
\label{eq:nphoton_sg}
\end{equation}
Here we assume that the decrease of $n$ due to photon absorptions 
by target atoms or molecules is negligible.
Because the number of photon $n$ is a constant, we can only determine
either $F_0$ or $\Delta$. Thus, the SGSP, Eq.~(\ref{eq:sgsp3d}), 
is characterized by only one of them. We use
the experimentally determined focal spot size $\Delta^2$ in this work, 
so $F_0$ is the parameter to be optimized. One may have the impression that
an accurately measured focal area is 
a prerequisite to perform calibrations
using \textsc{xcalib}. However, this is not the case.
One can easily show that the focal area $\Delta^2$ is factored out by changing 
integration variables, $x=x^\prime\Delta$ and $y=y^\prime\Delta$, in 
Eq.~(\ref{eq:absyield}). It is then found that the ion yields $Y_{\rm theo}^{(+q)}({\bf P})$, 
Eq.~(\ref{eq:absyield}), and correspondingly the normalization constant
$N(\bf P)$, Eq.~(\ref{eq:norm}), are proportional to the focal area. 
Hence, the dependency on the focal area for the calibrated CSDs, 
$y_\text{theo}^{(+q)}({\bf P})$ in Eq.~(\ref{eq:ionyield}), 
is canceled out after the normalization.

Next, we address the transmission of the X-ray optics. 
The energy delivered to the focal spot is given by the product of the pulse energy $E$ as an input parameter and the transmission $T$, namely $TE$, which is equivalent to the quantity of $n\omega$, where the X-ray photon energy $\omega$ is an input parameter. 
Equating them, the transmission $T$ is given by
\begin{equation}
T=\frac{n\omega}{E}.
\label{eq:ntot}
\end{equation}
Substituting Eq.~(\ref{eq:nphoton_sg}) into Eq.~(\ref{eq:ntot}),
the transmission $T$ for the SGSP is given by
\begin{equation}
T=\frac{\Delta^2\omega}{aE}F_0.
\label{eq:trans_sg}
\end{equation} 
If $\Delta$ is not known experimentally, then we can only determine the ratio
\begin{equation}\label{eq:ratio_sg}
\frac{T}{\Delta^2}=\frac{\omega F_0}{aE}.
\end{equation}
The length over which ions are collected along the X-ray beam is often, but not always, determined by a slit aperture in the spectrometer (see $\pm z_0/2$ marked at the top of Fig.~\ref{fig:beam_geo}).
When the Rayleigh range is much wider than the ion detector slit size or the molecular beam size, 
the $z$-dependence of the X-ray beam width is no more relevant.
If this is the case, the volume integration
in three dimensions in Eq.~(\ref{eq:absyield}) can be approximated to that in two dimensions in 
the $xy$-plane, assuming that $\Delta_1(z)=\Delta$ [see Eq.~(\ref{eq:sgsp2d}) for this case].

\subsection{Double Gaussian spatial profile}
Another spatial fluence profile employed in this work is 
a double Gaussian spatial profile (DGSP) consisting of 
a narrow, high main peak and a wide, low-fluence tail.
This profile consists of two Gaussian profiles. The first Gaussian profile given 
by Eq.~(\ref{eq:sgsp3d}) is characterized by the peak fluence $F_0$.
In addition to the peak fluence $F_0$, the second Gaussian profile is characterized
by two supplementary parameters: a fluence ratio $f_r$ and a width ratio $w_r$
between the first and second Gaussian profiles at $z=0$. 
The peak fluence and the beam size of the second Gaussian profile at $z=0$ are 
given by $f_rF_0$ and $w_r\Delta$, respectively.
Then the spatial profile is given by
\begin{equation}
\label{eq:dgsp3d}
F({\bf r};{\bf P})
=\frac{\Delta^2}{\Delta^2_{1}(z)}F_0e^{-\pi a\frac{x^2+y^2}{\Delta^2_{1}(z)}}
+\frac{(w_r\Delta)^2}{\Delta^2_{2}(z)}f_rF_0e^{-\pi a\frac{x^2+y^2}{\Delta^2_{2}(z)}},
\end{equation}
where the beam size $\Delta_{2}(z)$ of the second Gaussian profile is given by
\begin{equation}
\Delta_{2}(z)=w_r\Delta\sqrt{1+\left(\frac{z}{z_{R2}}\right)^2}. 
\label{eq:bsize2}
\end{equation}
The quantity $z_{R2}$ represents the Rayleigh range of the second Gaussian 
profile given by
\begin{equation}
z_{R2}=\frac{1}{2\ln(2)}\frac{\pi (w_r\Delta)^2}{\lambda}=w_r^2z_{R1}.
\label{eq:zr2}
\end{equation}
The first and second Gaussian profiles in the DGSP 
enable modeling an XFEL beam with a narrow intense hot spot on a broad tail, as demonstrated in previous work \cite{murphy2014}. 
It was also reported that an experimental spatial profile of an XFEL pulse may show one ideal peak as well as several additional peaks due to aberrations of the focus~\cite{nagler2017}, which could be modeled as a broad, low-fluence tail.
We thus limit the range of the values of $w_r$ and $f_r$ such that $w_r>1$ and $f_r<1$ 
so that the second Gaussian profile corresponds to a wide and low fluence tail. 
In case the optimization obtains a solution such that $w_r<1$ and $f_r>1$, 
the solution can be converted into an equivalent solution so that the first
Gaussian becomes narrower and higher (see Appendix \ref{sec:convert_sol}).

Next, we derive an expression for the transmission for the DGSP. 
Summing up the maximum amplitudes of the first and
the second Gaussian profiles, we introduce the global peak fluence $F_G$ given by
\begin{equation}
F_G=(1+f_r)F_0.
\label{eq:FG}
\end{equation}
Using the global peak fluence $F_G$ as well as $w_r$ and $f_r$, 
we characterize the DGSP, Eq.~(\ref{eq:dgsp3d}), by the set of three parameters ${\bf P}$ given by
\begin{equation}
\label{eq:dg_params}
{\bf P}=\left(F_G, w_r, f_r \right).
\end{equation}
Let $n_1$ and $n_2$ be a number of photons in the first and second Gaussian profiles 
in Eq.~(\ref{eq:dgsp3d}), respectively. Then these are given by
\begin{subequations}
\label{eq:n1n2_1}
\begin{eqnarray}
\label{eq:n1_1}
n_{1}&=&\frac{\Delta^2}{a}\frac{F_G}{1+f_r},\\
\label{eq:n2_1}
n_{2}&=&f_rw_r^2n_1,
\end{eqnarray}
The total number of photons $n$ is 
\begin{equation}
n=n_1+n_2=\frac{\Delta^2}{a}\frac{1+w_r^2f_r}{1+f_r}F_G.
\label{eq:ntot_dg}
\end{equation}
Substituting Eq.~(\ref{eq:ntot_dg}) into Eq.~(\ref{eq:ntot}),
we obtain a formula for the transmission for the DGSP,
\begin{equation}
T=\frac{\omega\Delta^2}{aE}\frac{1+w_r^2f_r}{1+f_r}F_G.
\label{eq:trans_dg}
\end{equation}
\end{subequations}
Again, if $\Delta^2$ is not accurately known, we can only find the ratio
\begin{equation}
\frac{T}{\Delta^2}=\frac{\omega}{aE}\frac{1+w_r^2f_r}{1+f_r}F_G.
\label{eq:ratio_dg}
\end{equation}
Because the width ratio $w_r$ of the second Gaussian profile changes at each step of 
the optimization procedure,
we adopted a specific grid scheme to accurately calculate 
the volume integration (see Appendix \ref{sec:grid_scheme}). 
\begin{figure}
\includegraphics[scale=0.5]{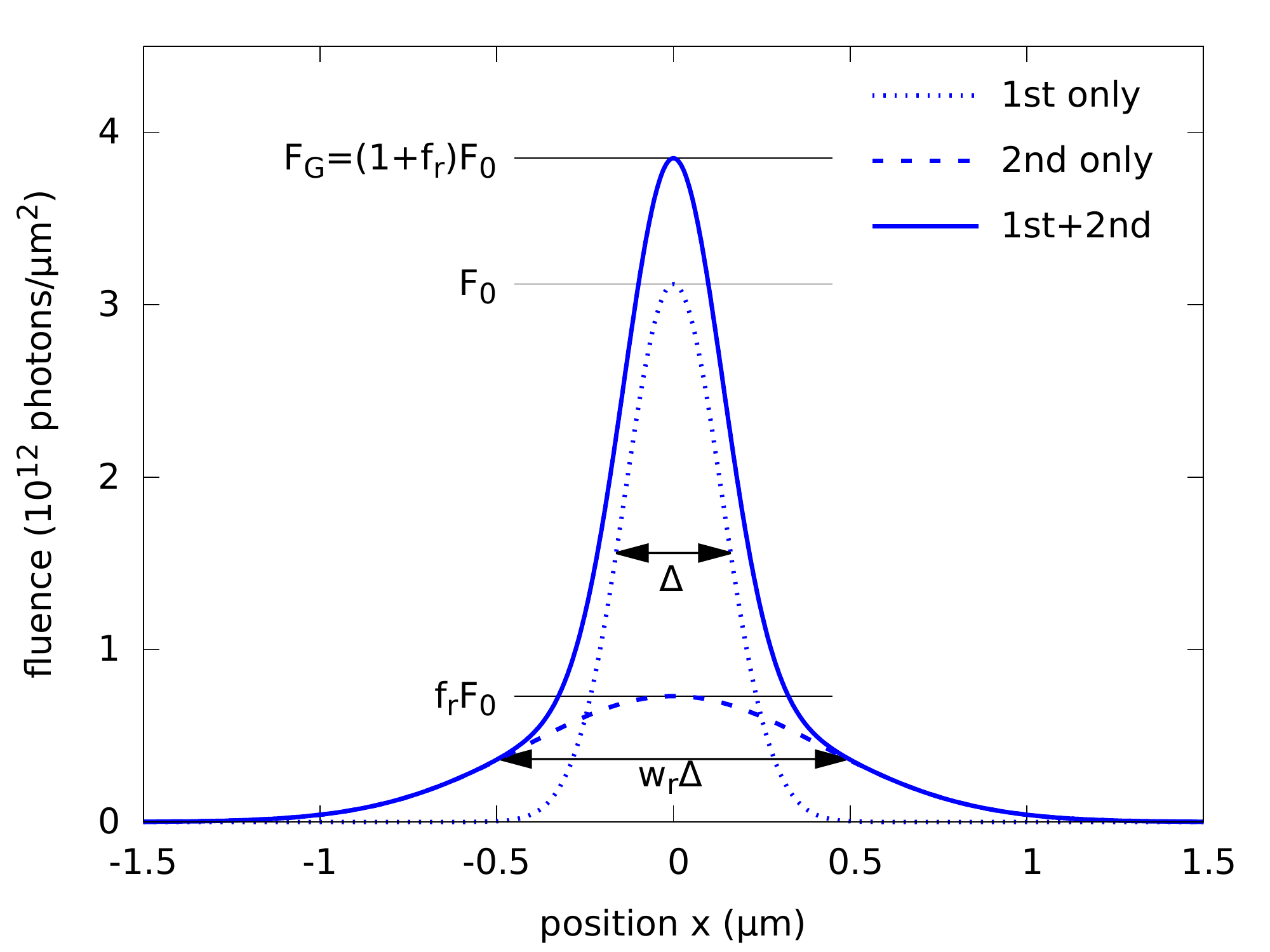}
\caption{\label{fig:beam_prof} The DGSP, Eq.~(\ref{eq:dgsp3d}), as a function of $x$ at $y=z=0$. 
The first and second Gaussian profiles are represented by the dotted and dashed lines, respectively.
The FWHM values $\Delta$ and $w_r\Delta$ and the peak fluence values $F_0$
and $f_rF_0$, and the global peak fluence $F_G$ are shown. 
The first Gaussian profile forms the narrow and high main peak, and the second Gaussian profile does the wide and low background. 
The values of parameters are given by Eq.~(\ref{eq:P6500eV_1st}). 
These values are obtained in the third example
in Sec.~\ref{sec:res_6500eV} calibrating the Ar CSD at 6.5 keV and 4.3 mJ.}
\end{figure}

A typical DGSP at $y=0$ and $z=0$ is shown in Fig.~\ref{fig:beam_prof}. It is seen in the figure
that the first and second Gaussian profiles form the narrow and high main peak and the wide 
and low fluence background, respectively. The set of three parameters ${\bf P}$ 
is given by $F_G=3.85\times 10^{12}$~photons/$\mu$m$^2$, $w_r=3.04$, and $f_r=0.234$. The set is obtained by
Ar calibration at 6.5 keV and 4.3 mJ demonstrated in Sec.~\ref{sec:res_6500eV} [see Eq.~(\ref{eq:P6500eV_1st})].
The beam sizes of the first and second Gaussian profiles, $\Delta_1(z)$ in Eq.~(\ref{eq:bsize1}) and $\Delta_2(z)$ in Eq.~(\ref{eq:bsize2}), are shown by
purple and green in Fig.~\ref{fig:beam_geo}. The Rayleigh ranges Eqs.~(\ref{eq:zr1}) 
and (\ref{eq:zr2}) and a detector of slit length $z_0$ are also shown 
in this figure. 

\subsection{Ion yields before and after volume integration}
Next, we show the calculated yields of Ar ions before and after the volume integration, 
Eq.~(\ref{eq:ionyield}), using the DGSP, Eq.~(\ref{eq:dgsp3d}), for illustrative purpose. 
We use \textsc{xatom}~\cite{xatom2018} to simulate ionization dynamics of isolated atoms 
interacting with intense XFEL pulses and calculate CSDs for a given fluence~\cite{son2011,jurek2016}.
The set of parameters ${\bf P}$ for the DGSP is the same as used for 
Figs.~\ref{fig:beam_geo} and \ref{fig:beam_prof}.

\begin{figure}
\includegraphics[scale=0.40]{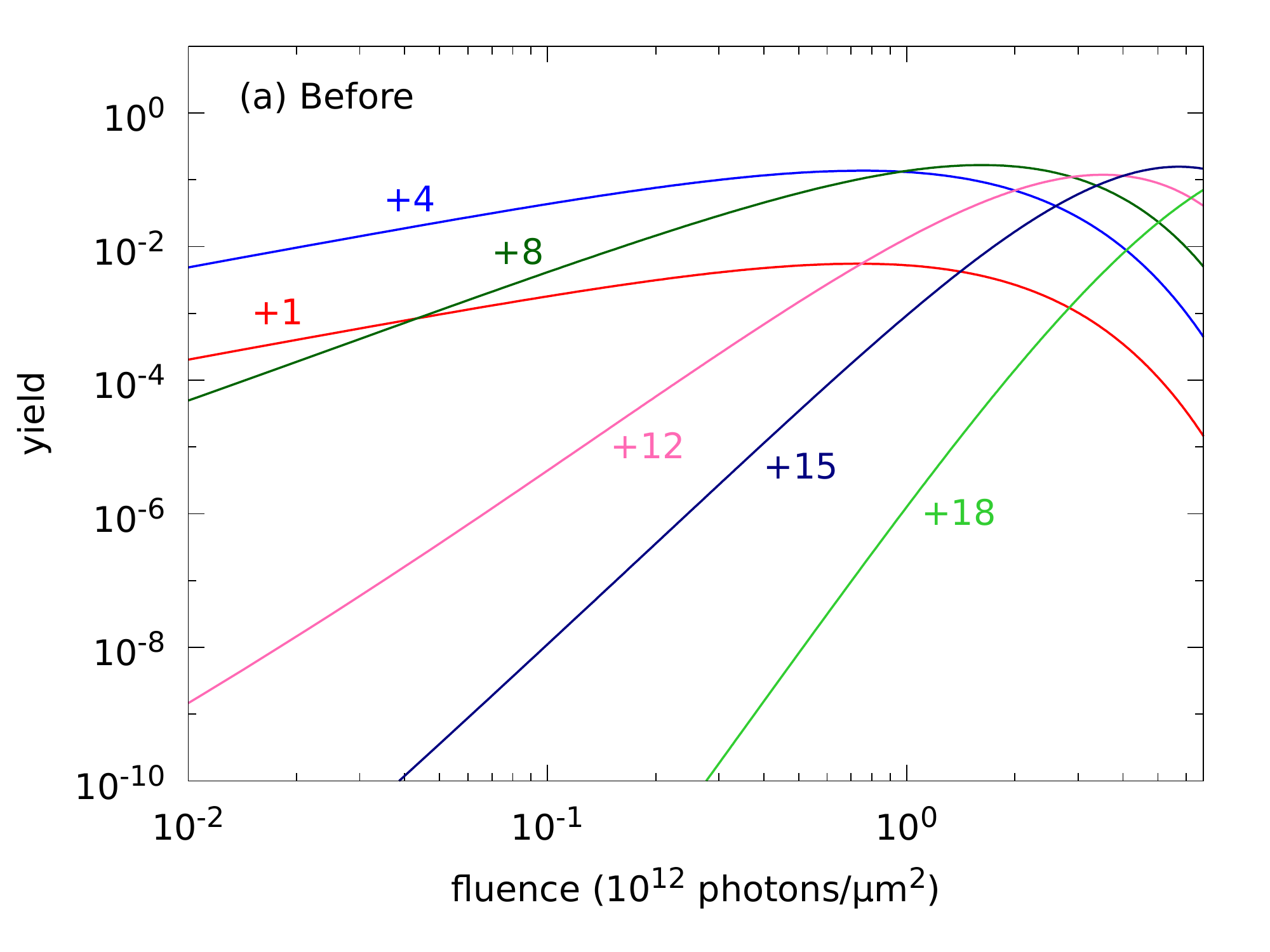}
\includegraphics[scale=0.40]{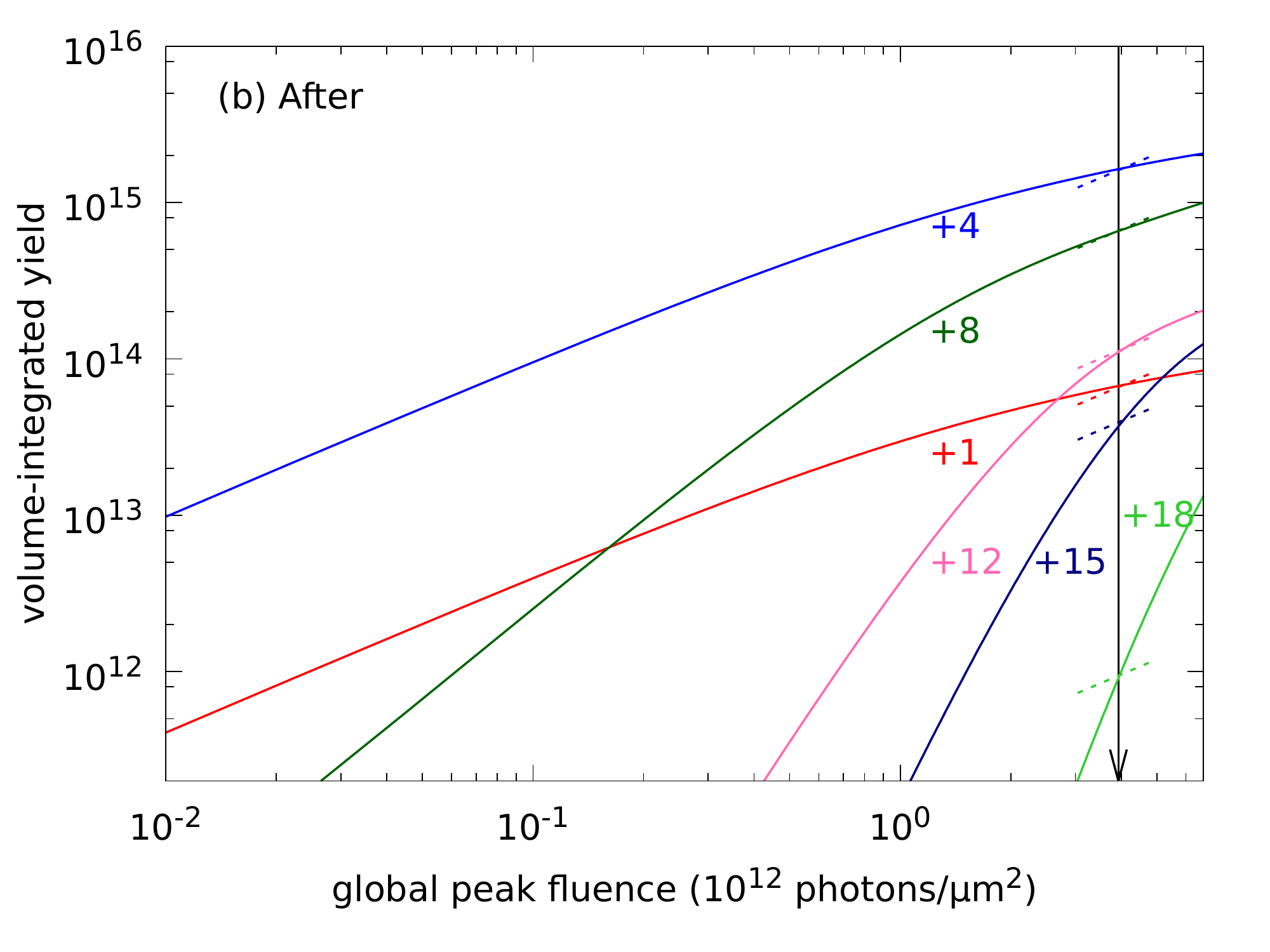}
\caption{\label{fig:Ar_yield} (a) Ion yields for several Ar charge states at 6.5 keV 
as a function of fluence before the volume integration, Eq.~(\ref{eq:absyield}).  
Note that we employ logarithmic scales for both axes.
(b) The same ion yields but after the volume integration using the calibrated parameters in Eq.~(\ref{eq:P6500eV_1st}).
The vertical arrow indicates $F_G=3.85\times 10^{12}$~photons/$\mu$m$^2$, which is the calibrated global peak fluence.
The dashed lines representing the slope of unity are marked at the vertical arrow to guide the eye.}
\end{figure}

The photon energy used in the simulations is 6.5~keV.
The calculated CSDs are shown in Fig.~\ref{fig:Ar_yield} as a function of the fluence. 
In Fig.~\ref{fig:Ar_yield}(a), the yields of Ar ions $\mathcal{Y}^{(+q)}_{\rm theo}(F)$, the integrand in Eq.~(\ref{eq:absyield}), as a function of fluence are shown for several charge states.
It is seen in the figure that, except for +18 which is the highest charge state of Ar, the ion yields $\mathcal{Y}^{(+q)}_{\rm theo}(F)$ reach a maximum at a certain fluence value and then they start decreasing due to target depletion, i.e., they are saturated.
Figure~\ref{fig:Ar_yield}(b) shows the volume-integrated absolute ion yields defined in Eq.~(\ref{eq:absyield}) as a function of global peak fluence $F_G$.
A vertical arrow indicates the global fluence value of $F_G=3.85\times 10^{12}$~photons/$\mu$m$^2$ in the set of ${\bf P}$, Eq.~(\ref{eq:P6500eV_1st}).
In Fig.~\ref{fig:Ar_yield}(b), the volume-integrated ion yields become flattened or saturated as $F_G$ increases, because of low-fluence contributions to their ion yields.

We utilize the slope of ion yields $s(F_G)$ as a function of global peak fluence $F_G$ to measure the degree of saturation,
\begin{equation}
s(F_G)=
\frac{\ln Y^{(+q)}_{\rm theo}(F_G+\Delta F_G)-\ln Y^{(+q)}_{\rm theo}(F_G)}{\ln(F_G+\Delta F_G)-\ln F_G},
\label{eq:slope}
\end{equation}
where the quantity $Y^{(+q)}_{\rm theo}(F_G)$ represent the absolute ion yield given by Eq.~(\ref{eq:absyield}). 
The other parameters in ${\bf P}$ are not shown for simplicity.
Note that we define the slope using a double logarithmic scale so that in the low fluence limit the value corresponds to the number of required photons to produce a certain charge state (see Appendix \ref{sec:slope}).
We regard a charge state as saturated at a given fluence $F_G$ if the slope of the associated ion yield is lowered by one or more in comparison with the low-fluence limit.
If the slope becomes even less than unity, especially for low charge states, the ion yield might be less sensitive to fluence values around the saturation point, and then it could introduce an unnecessary ambiguity into the optimization of the $F_G$ value.
Thus, it is conceivable to select charge states based on the slope of their volume-integrated ion yields.
The pieces of dashed lines in Fig.~\ref{fig:Ar_yield}(b) are marked to guide the eye, representing the slope of unity to be compared with the calculated slope of ion yields at the calibrated fluence $F_G$.
It is found that the ion yields for $q \leq +9$ are saturated and their slope is less than unity ($+1$, $+4$, and $+8$ are shown in the figure) before reaching $F_G=3.85\times 10^{12}$~photons/$\mu$m$^2$. 
We attempt to remove these charge states in the optimization, which will be demonstrated in Sec.~\ref{sec:res_6500eV}.
Note that the saturation effect needs to be carefully taken into consideration, especially when calibration is performed at very high fluences.
If the majority of charge states are saturated and their ion yields become insensitive to different fluence values, the result of the calibration procedure becomes unreliable.

\subsection{Optimization}
\begin{figure}
\includegraphics[scale=0.5]{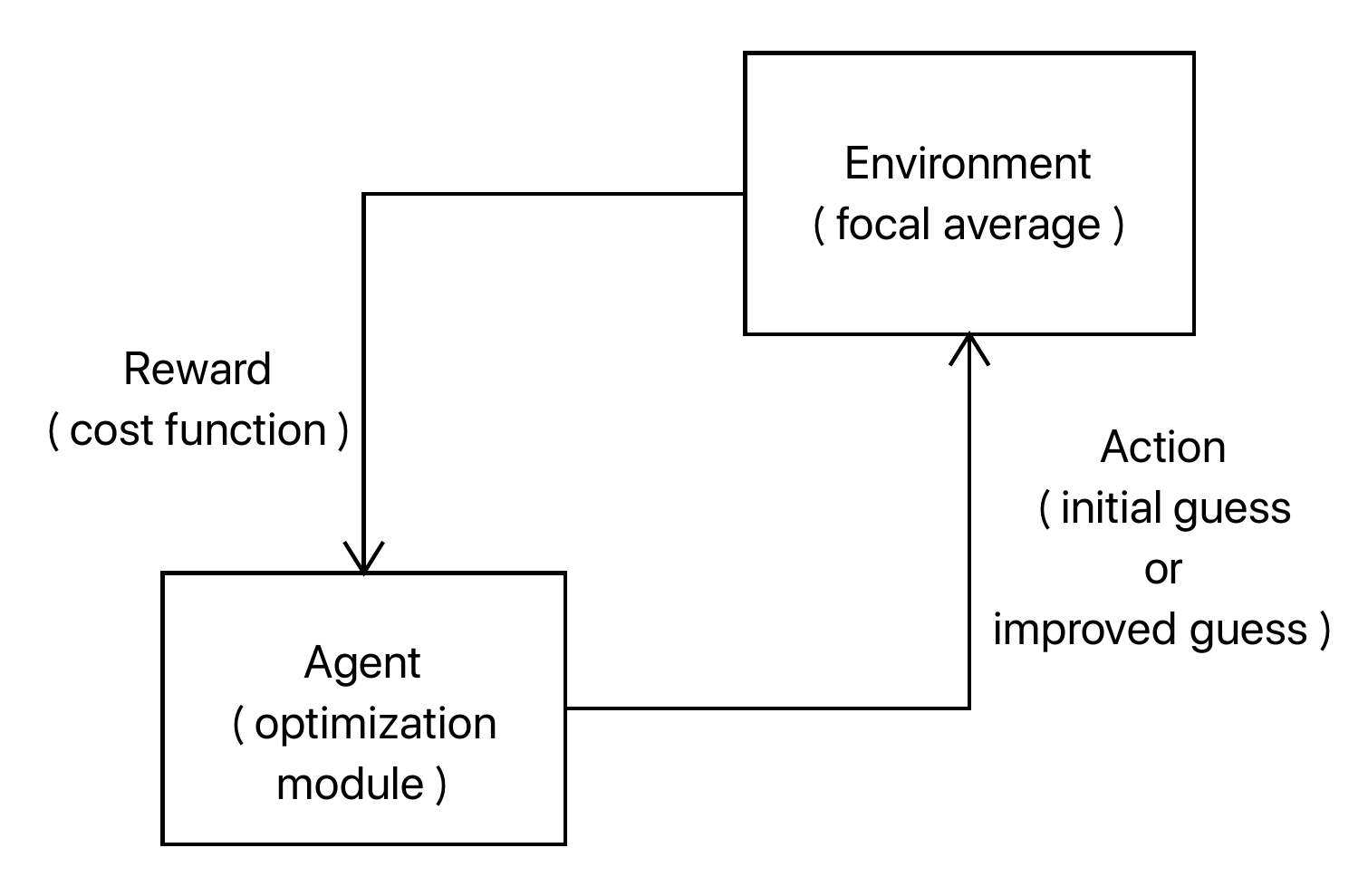}
\caption{\label{fig:reinforce} The concept of the reinforce learning in machine learning \cite{raschka2015}.}
\end{figure}

Here we explain the automated calibration procedure for spatial fluence profiles
employing the volume integration scheme established in the previous subsections.
The aim of calibration is to find out the best parameter set ${\bf P}$ 
that minimizes the difference between the volume integrated theoretical result
and the experimental result. In previous works \cite{rudek2012,fukuzawa2013,murphy2014,rudenko2017}, 
the calibrations were conducted by manually exploring the parameter space.
The manual exploration is inefficient because the direction to a solution
from an initial guess is not known in general. Hence, the manual procedure
becomes impractical when the dimension of ${\bf P}$ grows
or a number of calibrations must be performed. 
In \textsc{xcalib} the calibration is performed by reinforcement learning 
\cite{raschka2015} combined with optimization modules in \textsc{python} 
that automatize exploring the parameter space. The concept of the reinforcement 
learning is sketched in Fig.~\ref{fig:reinforce}. In this figure, the Agent 
gives the Environment Action, then the Environment returns the Reward. 
The Agent repeats the procedure to find the Action that minimizes 
or maximizes the Reward in a trial-and-error approach. In \textsc{xcalib}, 
the Agent is built up employing optimization modules in \textsc{python}. 
The Action and the Environment correspond to an initial (improved) guess 
and the calculation of the volume integration using the guess. 
The Reward in Fig.~\ref{fig:reinforce} corresponds to 
a cost function $\delta(\bf P)$ defined by the difference between
the volume-integrated theoretical result and the experimental data.
Since the next direction in the parameter space from a current guess
is determined by established mathematical algorithms, 
the reinforcement learning approach is much more efficient than
the manual procedure. Although we calibrate at most three parameters
in this work, using \text{xcalib} will be critical when
calibrating the fluence profiles in high-dimensional parameter 
spaces.

In this work, we define the cost function by the sum of quadratic-logarithmic  
differences between the theoretical $y^{(+q)}_{\rm theo}({\bf P})$ 
and the experimental ion yield $y^{(+q)}_{\rm expt}$, 
\begin{equation}
\delta({\bf P})=\sum_{{\rm selected}~q} 
\left( \log\frac{y^{(+q)}_{\rm theo}(\bf P)}{y^{(+q)}_{\rm expt}} \right)^2.
\label{eq:cost_func}
\end{equation}
The experimental ion yields are normalized in the same manner as the theoretical 
result [see Eqs.~(\ref{eq:ionyield}) and (\ref{eq:norm})].
In Eq.~(\ref{eq:cost_func}), the summation runs over selected charge states. 
The set of selected charge states is formed by removing the charge states whose ion yields are not 
accurate enough in theory and/or experiment.
At the end of each iteration in the optimization, 
the ion yield is normalized according to Eqs.~(\ref{eq:ionyield})
and (\ref{eq:norm}). 
Although some charge states are excluded in the sum of Eq.~(\ref{eq:cost_func}), an optimized result depends on them through the normalization, Eq.~(\ref{eq:norm}).
We keep them in the normalization condition for compatibility with previous works \cite{rudek2012, fukuzawa2013,murphy2014}.
The advantage of using a logarithmic function is that the relative weights of the ion yields over the wide range of charge states become comparable in amplitude. 
Therefore, tiny ion yields for high charge states and large ion yields for other charge states can be treated on equal footing in 
the optimization.

\section{Results}
\label{sec:res}
In this section, we demonstrate the \textsc{xcalib} toolkit for three examples.
The first example is the Ar calibration at 5.5 keV for the xenon (Xe) 
experiment performed at SACLA \cite{fukuzawa2013}.
We confirm that the results produced by \textsc{xcalib} are consistent 
with those in \cite{fukuzawa2013}. The second example is the Ar calibration at 805~eV 
for the C$_{60}$ experiment at LCLS \cite{murphy2014}. We demonstrate that 
the second Gaussian profile in the DGSP, Eq.~(\ref{eq:dgsp3d}), is essential to model
the low fluence tail of an XFEL pulse, as previously shown in 
\cite{murphy2014,nagler2017}.
In the third example, we study the effect of an attenuator on the 
spatial fluence distribution by performing an Ar calibration at $6.5$ keV 
for a recent experiment \cite{rudek2018}.

\subsection{Ar calibration with a single Gaussian spatial profile}
\label{sec:res_5500eV}

In the SACLA experiment \cite{fukuzawa2013}, the photon energy and the pulse energy were $5.5$~keV and $239~\mu$J, respectively.
The nominal focal spot area was $\Delta^2=1 \times 1$~$\mu$m$^2$.
The molecular beam size was about 2~mm.
The peak fluence $F_0$ of Eq.~(\ref{eq:sgsp3d}) was manually calibrated 
to reproduce the experimental ratio 
\begin{equation}
\frac{y^{(+8)}_{\rm expt}(F_0)+y^{(+9)}_{\rm expt}(F_0)}{y^{(+3)}_{\rm expt}(F_0)+y^{(+4)}_{\rm expt}(F_0)},
\label{eq:ratio}
\end{equation}
where the numerator represents the sum of ion yields of Ar$^{8+}$ and Ar$^{9+}$  
produced by two-photon absorption, and the denominator is that of Ar$^{3+}$ and Ar$^{4+}$ 
produced by one-photon absorption. The ratio thus gives us the relative contribution
between two-photon and one-photon absorption processes.
However, \textsc{xcalib} does not require such additional physical considerations 
on the number of required photons to produce a certain charge state 
in order to define a cost function.

The Rayleigh range calculated via Eq.~(\ref{eq:zr1}) with the given photon energy and the focal spot area is $z_{R1}=10$ mm, which is five times larger than the molecular beam size. 
In such a situation, it can be assumed that the atoms were subject to the same fluence value in the $z$-direction. 
Under this consideration, the focal size in $\Delta_{1}(z)$, Eq.~(\ref{eq:sgsp3d}),
is approximated by
$\Delta_{1}(z) \approx \Delta$.
The SGSP, Eq.~(\ref{eq:sgsp3d}), is then simplified to
\begin{equation}
F(x,y;F_0)=F_0e^{-\pi a\frac{x^2+y^2}{\Delta^2}}.
\label{eq:sgsp2d}
\end{equation}
In the following, we revisit the calibration for 
the SACLA experiment with \textsc{xcalib}, employing
the two-dimensional (2D) version of the SGSP, Eq.~(\ref{eq:sgsp2d}).

\begin{figure}
\includegraphics[width=0.45\textwidth]{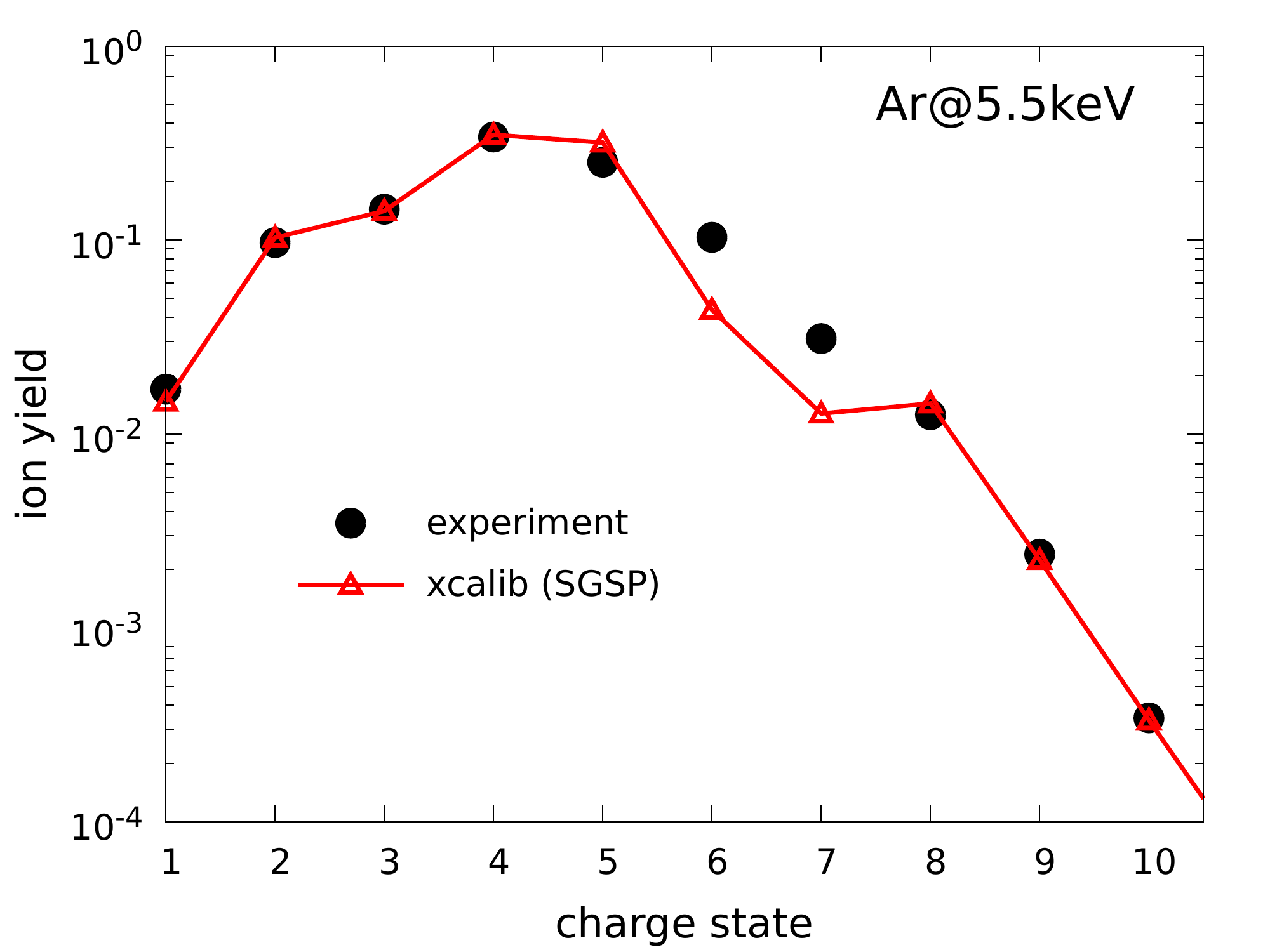}
\includegraphics[width=0.45\textwidth]{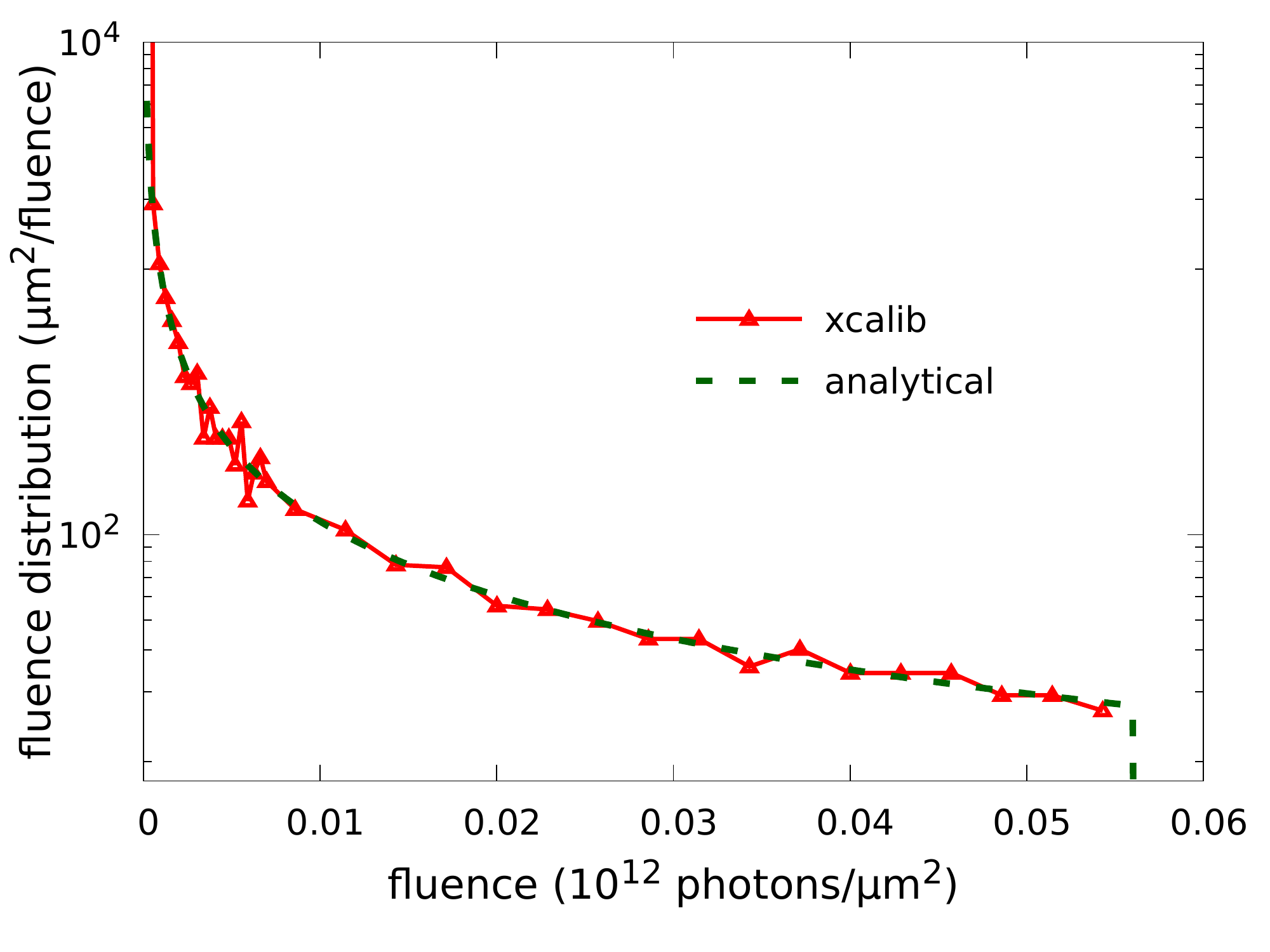}
\caption{\label{fig:Ar_sacla} (a) Ar CSD at 5.5 keV in \cite{fukuzawa2013} is shown.
The black line with dots shows the experimental results. The red line with triangles shows the 
theoretical results obtained by \textsc{xcalib} with the SGSP, Eq.~(\ref{eq:sgsp2d}).
(b) Numerical and analytical FDFs for the SGSP, Eq.~(\ref{eq:sgsp2d}). 
The analytical formula is given by Eq.~(\ref{eq:flu_dist_sg})
}
\end{figure}

The experimental Ar CSD is shown by circles (black) in Fig.~\ref{fig:Ar_sacla}(a).
The set of all charge states for Eqs.~(\ref{eq:ionyield}), (\ref{eq:norm}) and (\ref{eq:norm_cond})
consists of the charge states from $+1$ to $+10$ observed in the experiment. 
The set of selected charge states to calculate the cost function, Eq.~(\ref{eq:cost_func}), 
is obtained by removing the charges states of $+6$ and $+7$ from 
the set of all charge states whose ion yields are underestimated in theory  
\cite{fukuzawa2013,murphy2014,rudenko2017}. 
To obtain the optimized value of $F_0$, 30 \textsc{xcalib} runs were submitted with
the initial guesses of the peak fluence $F_0$ equidistantly distributed over 
$F_0~(10^{12}~{\rm photons}/\mu {\rm m}^2) \in [0.036,0.36]$. The optimized peak fluence value is 
$F_0=0.056 \times 10^{12}$~photons$/\mu$m$^2$ corresponding to the lowest cost function 
value, Eq.~(\ref{eq:cost_func}). The transmission value of $23.2\%$ calculated using 
Eq.~(\ref{eq:trans_sg}) is consistent with 22.3\% obtained in \cite{fukuzawa2013}
by performing a three-dimensional
volume integration. Therefore the integration with respect 
to the $z$-axis does not affect the result in this case. 
The 2D volume-integrated Ar CSD is shown by triangles (red) in Fig.~\ref{fig:Ar_sacla}(a). 
Comparing with the experimental result, it is seen that the ion yield of $+5$ 
is slightly overestimated, while those of $+6$ and $+7$, excluded in the cost function,
are significantly underestimated.
A possible reason for this discrepancy is the neglect of higher-order many-electron corrections in our theoretical model.

We further examine \textsc{xcalib} by comparing numerical and analytical 
fluence distribution functions (FDFs).
The FDF is defined by the area per unit fluence occupied by a certain 
fluence value in a given spatial profile. For instance, the FDF vanishes 
at the peak fluence and goes to infinity at zero fluence.
A numerical FDF is obtained by making a histogram of fluence values 
multiplied by a constant. The constant is given by a unit volume made by spatial
grid points divided by the fluence bin size. An analytical formula
is obtained for the case of the SGSP, Eq.~(\ref{eq:sgsp2d}) 
(see appendix \ref{sec:flu_dist_deriv}). In Fig.~\ref{fig:Ar_sacla}(b),
the numerical and analytical FDFs are shown by solid (red) and dashed lines (green).
It is seen that both of them agree very well. The FDF vanishes above the peak 
fluence $F_0=0.056\times 10^{12}$~photons$/\mu$m$^2$. The noisy behavior in the numerical result
for fluence values comes from low statistics.

\subsection{Ar calibration with a double Gaussian spatial profile}
\label{sec:res_805eV}

\begin{figure}
\includegraphics[width=0.45\textwidth]{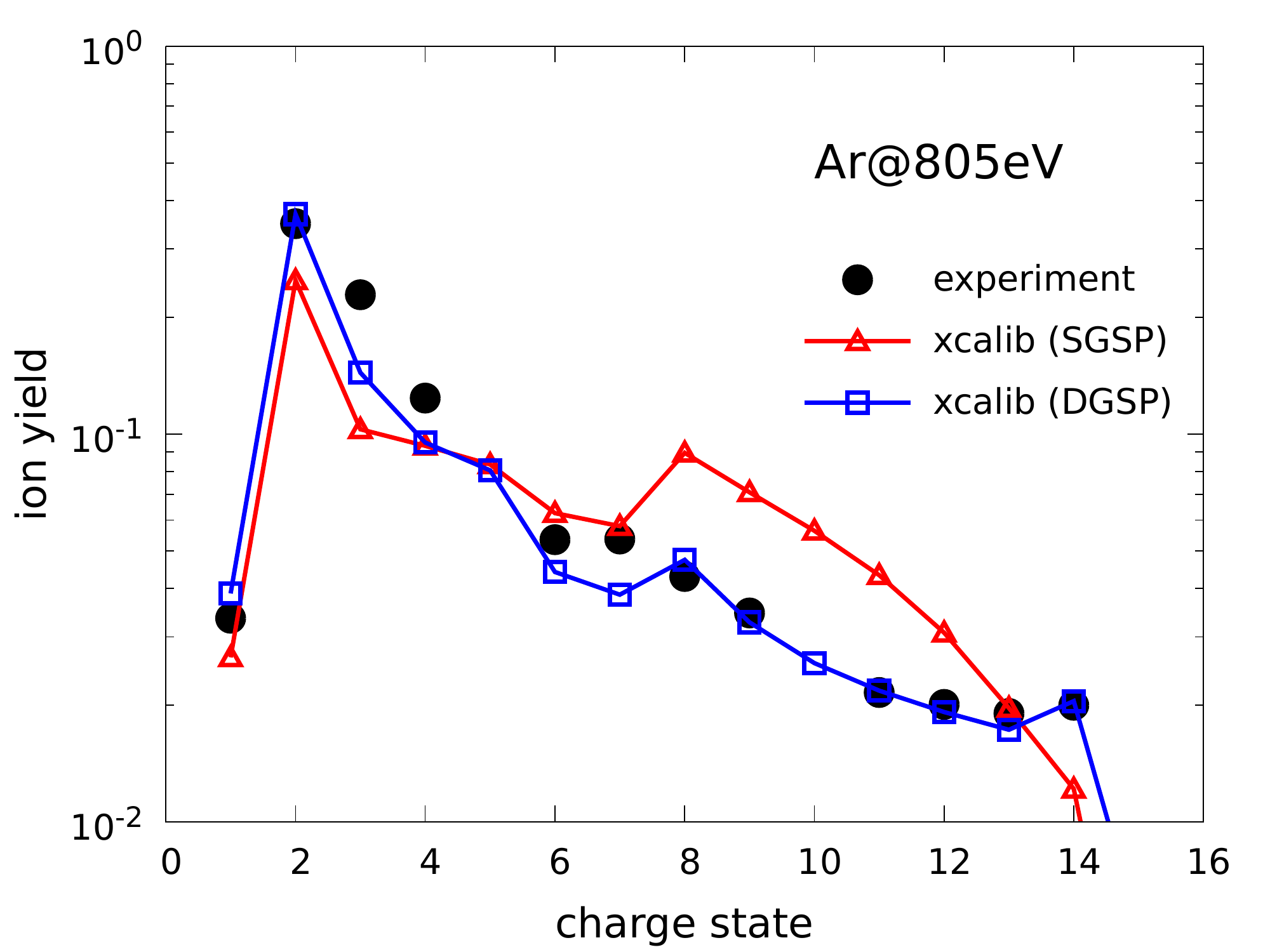}
\includegraphics[width=0.45\textwidth]{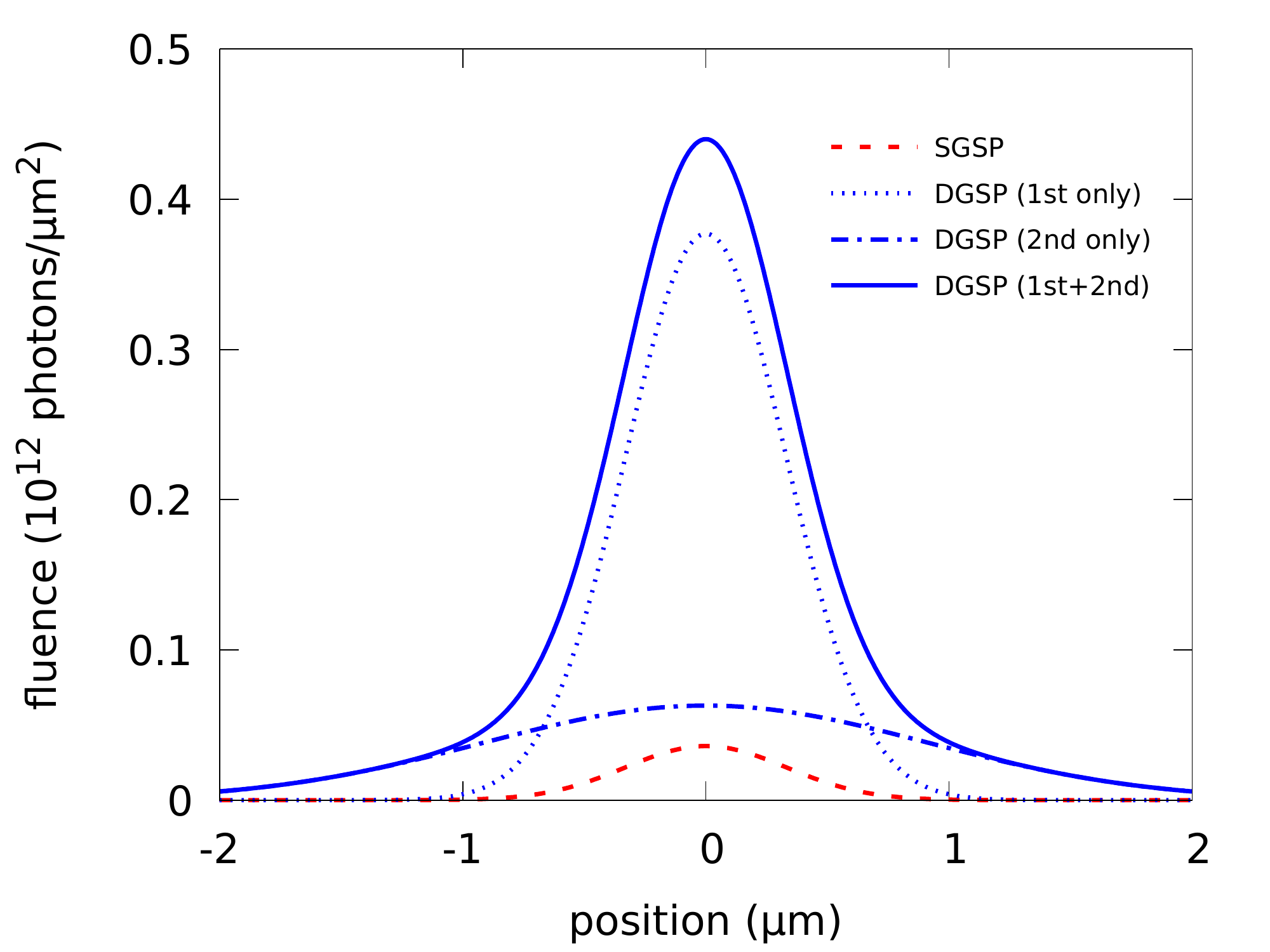}
\caption{\label{fig:Ar_C60} 
(a) Experimental Ar CSD at 805 eV obtained in \cite{murphy2014} is shown by black dots.
Triangles indicate the volume-integrated results with calibrated SGSP in Eq.~(\ref{eq:sgsp3d}), 
and squares are those with calibrated DGSP in Eq.~(\ref{eq:dgsp3d}) in this work. 
(b) Calibrated SGSP, Eq.~(\ref{eq:sgsp3d}), and DGSP, Eq.~(\ref{eq:dgsp3d})
at $y=0$. }
\end{figure}

Here we revisit the Ar calibration at $805$~eV performed for the C$_{60}$ experiment at LCLS \cite{murphy2014}. 
The pulse energy was $1.15$~mJ and the nominal focal area was $\Delta^2=1.38\times 1.38$~$\mu$m$^2$.
The slit size was 1.6~mm~\cite{osipov_private}.
The experimental Ar CSD is shown by black dots in Fig.~\ref{fig:Ar_C60}(a). 
The charge states of $+5$ and $+10$ are not shown because of experimental
uncertainty. Thus, when evaluating Eqs.~(\ref{eq:norm}) and (\ref{eq:norm_cond}),
the set of charge states consists of $+1$ to $+14$
without $+5$ and $+10$. 
The set of selected charge states to calculate the cost function, Eq.~(\ref{eq:cost_func}),
is constructed without the charge states of $+5$, $+6$, $+7$, and $+10$. 

The calibration was first attempted employing a SGSP, Eq.~(\ref{eq:sgsp3d}), 
taking the Rayleigh range into consideration. 
The photon energy is smaller than that used in Sec.~\ref{sec:res_5500eV}, so the Rayleigh range is smaller here.
The calculated value is $z_{R1}=2.8$~mm, which is now comparable with the slit size.
Therefore, we use the three-dimensional integration in Eq.~(\ref{eq:ionyield}) 
with the three-dimensional SGSP in Eq.~(\ref{eq:sgsp3d}), 
in order to obtain the CSDs.
The result is shown using triangles (red) in Fig.~\ref{fig:Ar_C60}(a). The calibration clearly 
fails to reproduce the experimental result. 
The ion yields for low charge states are underestimated, 
whereas those for middle to high charge states are overestimated, as was also found in \cite{murphy2014}. 
This failure implies that the beam profile is more complex than a SGSP,
so a DGSP was introduced in \cite{murphy2014} to overcome this problem. 

Here we also employ a DGSP, Eq.~(\ref{eq:dgsp3d}).
27 initial guesses uniformly distributed in the parameter range of
\begin{equation}
F_G~(10^{12}~{\rm photons}/\mu{\rm m}^2)\in [0.357,1.79],~w_r \in [1.1,3.1],~f_r \in [0.1,0.5],
\end{equation}
were employed to perform the optimization.
The solution giving the lowest local cost function value for the set of  
selected charge states is given by ${\bf P}=(F_G, w_r, f_r)=(0.44 \times 10^{12}~{\rm photons}/\mu {\rm m}^2, 2.77, 0.167)$.
The result is given by squares (blue) in Fig.~\ref{fig:Ar_C60}(a).
The agreement between the result of \textsc{xcalib} and the experimental results 
is now much better.
The low fluence wing supported by the second Gaussian profile in Eq.~(\ref{eq:dgsp3d}) enhances
the ion yields of low charge states.
The DGSP, Eq.~(\ref{eq:dgsp3d}), with the calibrated parameter set
at $y=0$ is depicted in Fig.~\ref{fig:Ar_C60}(b) together with the SGSP, Eq.~(\ref{eq:sgsp3d}). 

\subsection{Ar calibration with attenuated beams}
\label{sec:res_6500eV}
In this demonstration, we calibrate the three data sets of Ar CSDs at 6.5 keV 
taken for a recent experiment on Xe \cite{rudek2018}. 
The first data set consists of Ar CSDs measured without an attenuator. 
The pulse energy fluctuates from shot to shot, so the Ar CSD data are 
binned according to the pulse energies of 4.5, 4.3, and 4.1~mJ.
A silicon attenuator was used to filter 58\% of the full-power beam, 
providing the pulse energies of 2.61, 2.49, and 2.38~mJ (the second data set), 
and 20\% corresponding to 0.90, 0.86, and 0.82~mJ (the third data set). 
We calibrate these three data sets (100\%, 58\%, and 20\%), as listed in Table~\ref{tab:att}, to study the effect of the
attenuator on the spatial fluence profile. It has been believed  
that an attenuator would not change the spatial profile of XFEL pulses, 
but no comprehensive studies have been reported so far on this subject.

\begin{table}
\caption{\label{tab:att} Calibrated DGSPs, Eq.~(\ref{eq:dgsp3d}), 
for non-attenuated (100\%) and attenuated (58\% and 20\%) beams at 6.5 keV. 
These parameters were extracted from the experimental Ar CSDs of \cite{rudek2018}. } 
\begin{ruledtabular}
\begin{tabular}{ccrrrrr}
& \shortstack{$E$ \\ (mJ)} & 
\shortstack{$F_G$ \\ (photons/$\mu$m$^2$)} & 
\shortstack{$w_r$ \\ \ } & 
\shortstack{$f_r$ \\ \ } & 
\shortstack{$w_r^2 f_r$ \\ \ } & 
\shortstack{{$T/\Delta^2$} \\ ($\mu$m$^{-2}$)} \\
\hline
\multirow{3}{*}{\shortstack{non-attenuated \\ (100\%)}} 
& 4.5\phantom{0} &   $4.01\times10^{12}$  &   2.91   & 0.266  & 2.25 & 2.70 \\
& 4.3\phantom{0} &   $3.93\times10^{12}$  &   3.10   & 0.281  & 2.70 & 3.12 \\
& 4.1\phantom{0} &   $3.90\times10^{12}$  &   3.29   & 0.275  & 2.98 & 3.50 \\
\hline
\multirow{3}{*}{\shortstack{attenuated \\ (58\%)}}
& 2.61           &   $2.95\times10^{12}$  &   3.69   & 0.160  & 2.18 & 3.65 \\
& 2.49           &   $3.02\times10^{12}$  &   3.98   & 0.168  & 2.66 & 4.49 \\
& 2.38           &   $2.95\times10^{12}$  &   4.16   & 0.171  & 2.96 & 4.95 \\
\hline
\multirow{3}{*}{\shortstack{attenuated \\ (20\%)}}
& 0.90           &   $1.27\times10^{12}$  &   5.37   & 0.135  & 3.89 & 7.18  \\
& 0.86           &   $1.44\times10^{12}$  &   6.02   & 0.172  & 6.23 & 12.19  \\
& 0.82           &   $1.39\times10^{12}$  &   6.30   & 0.175  & 6.95 & 13.53  
\end{tabular}
\end{ruledtabular}
\end{table}

The nominal focal spot area in the experiment was estimated as 
$\Delta^2=0.35\times 0.3$~$\mu$m$^2$ with an elliptic focal shape. 
In the numerical method described in Sec.~\ref{sec:num}, a circularly shaped 
focal area is assumed. We numerically confirmed that the volume-integrated Ar CSD and the FDF
at $4.3$~mJ do not change noticeably when an elliptic focal shape is explicitly used.
Hence, we keep using a circularly shaped focal spot, 
i.e., $\Delta^2=0.324\times0.324$~$\mu$m$^2$, in the following. 
The focal size is much smaller than that used in Sec.~\ref{sec:res_5500eV}, 
so the Rayleigh range is smaller.
It is calculated as $z_{R1}=1.25$~mm, which is quite comparable 
with the slit size used in experiment ($z_0=1.0$~mm)~\cite{rudek2018}. 
Therefore, we start with three-dimensional (3D) volume integration, but two-dimensional (2D) volume integration will also be tested later on.
Here we assume a DGSP, because a DGSP is more general than a SGSP.
We explain the numerical procedure for the pulse energy of $4.3$~mJ in the following. 
The same procedure is applied for the other pulse energies. 

\begin{figure}
\includegraphics[width=0.70\textwidth]{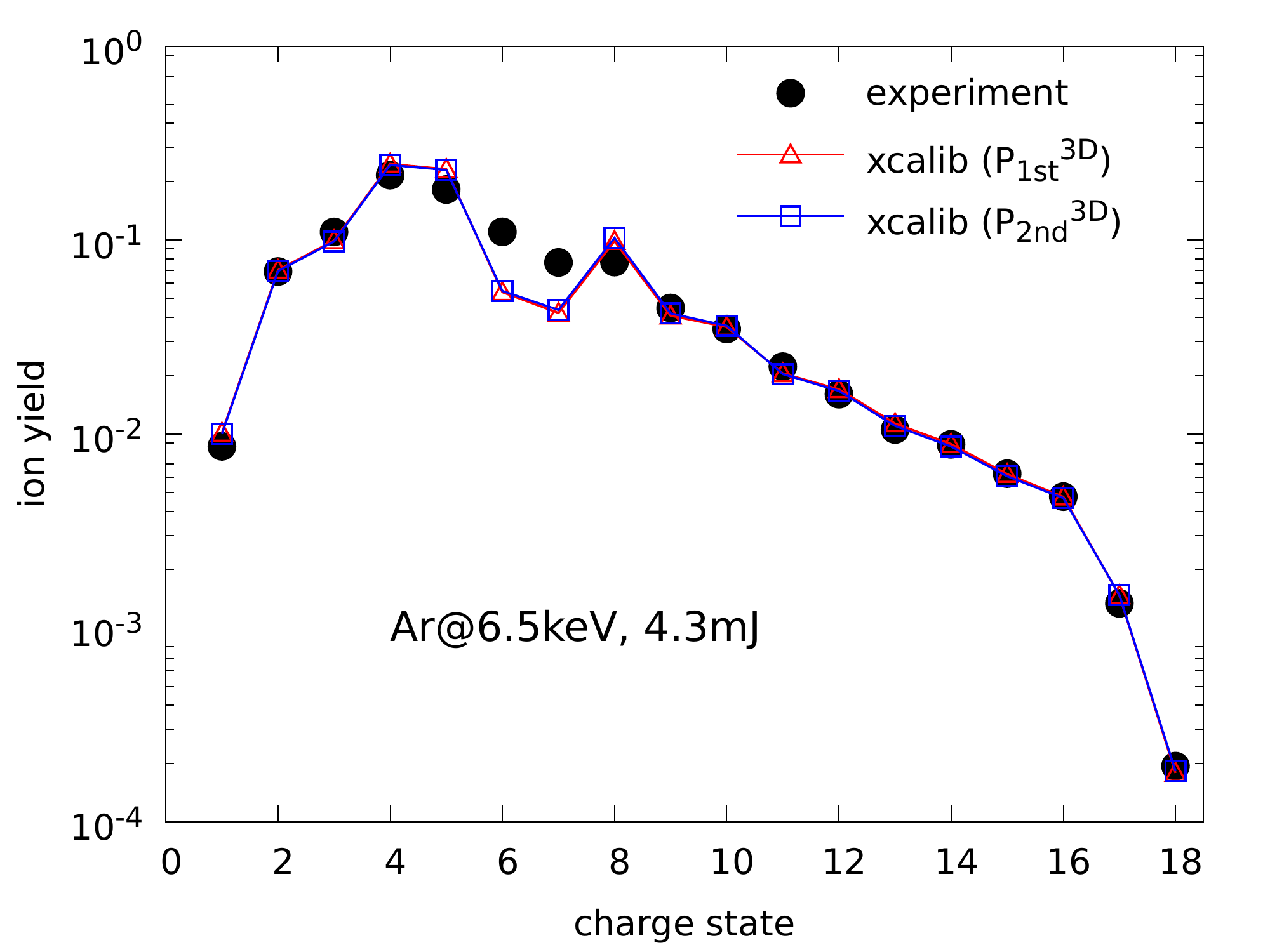}
\caption{\label{fig:screening} Ar CSDs at 6.5~keV and 4.3~mJ.
Circles (black) represent experimental result. Triangles (red) and squares (blue) 
represent the solutions ${\bf P}_{\rm 1st}^{\rm 3D}$, Eq.~(\ref{eq:P6500eV_1st}),
 and ${\bf P}_{\rm 2nd}^{\rm 3D}$, Eq.~(\ref{eq:P6500eV_2nd}), respectively, obtained by \textsc{xcalib}.
These are obtained by including or excluding some charge states whose volume-integrated ion yields are saturated.}
\end{figure}

The set of all charge states ranges from $+1$ to $+18$. The set of selected charge
states is obtained by removing $+6$ and $+7$, whose ion yields are inaccurate in theory. 
27 initial guesses were uniformly 
distributed in the parameter range given by
\begin{equation}
F_G~(10^{12}~{\rm photons}/\mu{\rm m}^2) \in [0.357, 1.79],~w_r \in [1.1,3.1],~f_r \in [0.1,0.5].
\end{equation}
The solution corresponding to the lowest cost function value, Eq.~(\ref{eq:cost_func}),
is given by
\begin{equation}
{\bf P}_{\rm 1st}^{\rm 3D}=(F_G, w_r, f_r)=(3.85\times 10^{12}~{\rm photons}/\mu {\rm m}^2, 3.04, 0.234).
\label{eq:P6500eV_1st}
\end{equation}
The beam geometry of the calibrated DGSP, Eq.~(\ref{eq:dgsp3d}), for the set of parameters 
${\bf P}_{\rm 1st}^{\rm 3D}$ was shown in Fig.~\ref{fig:beam_geo}. In Fig.~\ref{fig:Ar_yield}(b),
the volume-integrated ion yields, Eq.~(\ref{eq:ionyield}), for several charge states
for the set ${\bf P}_{\rm 1st}^{\rm 3D}$ were shown. The ion yields were 
calculated by changing $F_G$ values 
with $w_r$ and $f_r$ values being fixed. 
As discussed earlier, because of the saturation effect, the ion yields of low charge states are less sensitive to the change of the global fluence value around $F_G=3.85\times 10^{12}$~photons/$\mu$m$^2$ in optimization, which might give rise to an ambiguity in the determination of the global fluence value. 
Specifically, the ion yields of the charge states from +1 to +9 have a slope that is less than unity.
Therefore, a second optimization was performed removing these charge states from the set of charge states selected for the optimization procedure. 
Then we obtained the optimized set of parameters given by
\begin{equation}
{\bf P}_{\rm 2nd}^{\rm 3D}=(F_G, w_r, f_r)=(3.93 \times 10^{12}~{\rm photons}/\mu {\rm m}^2, 3.10, 0.282).
\label{eq:P6500eV_2nd}
\end{equation}
These two different optimizations using different sets of charge states give rather similar parameters.
Also both methods, ${\bf P}_{\rm 1st}^{\rm 3D}$ (red triangles) and ${\bf P}_{\rm 2nd}^{\rm 3D}$ (blue squares), provide almost the same Ar CSDs, as illustrated in Fig.~\ref{fig:screening}.
In the following applications, we keep the charge selection procedure using the slope of volume-integrated ion yields.
The transmission calculated using Eq.~(\ref{eq:trans_dg}) with ${\bf P}_{\rm 2nd}^{\rm 3D}$ is $32.8\%$.

Here we attempt to compare 2D and 3D volume integrations to perform the calibration. 
The result obtained via 2D volume integration after the charge selection is given by
\begin{equation}
{\bf P}_{\rm 2nd}^{\rm 2D}= (F_G, w_r, f_r)=(3.93\times 10^{12}~{\rm photons}/\mu {\rm m}^2, 3.10, 0.281),
\end{equation}
which practically coincides with the result of Eq.~(\ref{eq:P6500eV_2nd}). 
The transmission is calculated using Eq.~(\ref{eq:trans_dg}) as $32.6\%$, 
and the volume-integrated CSD with this calibrated parameter set looks quite similar 
to those in Fig.~\ref{fig:screening}.
The almost identical results of ${\bf P}_{\rm 2nd}^{\rm 3D}$ and ${\bf P}_{\rm 2nd}^{\rm 2D}$ 
indicate that the integration for the $z$-direction does not affect the final Ar CSDs. 
We confirm that this is also true for other pulse energies. Therefore,
we will show further results for ${\bf P}_{\rm 2nd}^{\rm 2D}$ only.

In Table~\ref{tab:att}, we list all numerical results for three data sets (9 different pulse energies).
In each row, the global peak fluence $F_G$, the width ratio $w_r$, and the fluence ratio $f_r$ are given. 
The energy ratio (or the ratio of the total numbers of photons) between the first and second Gaussians is calculated using Eq.~(\ref{eq:n2_1}), $n_2 / n_1 = w_r^2 f_r$.
The ratios $T/\Delta^2$ calculated using Eq.~(\ref{eq:ratio_dg}) are listed in the last column of the table, assuming that the input pulse energy $E$ is known.
When inspecting Table \ref{tab:att}, we find interesting trends. 
Our calibration suggests that $F_G$ decreases more slowly than $E$. 
Note that if the shape of the spatial fluence distribution were unchanged, $F_G$ would be exactly proportional to $E$.
Our calibrated parameters imply that the shape of the spatial fluence distribution of the X-ray beam appears to be affected by the attenuation process. 
This fact becomes even more evident when we consider the width ratio $w_r$, which increases monotonically with decreasing pulse energy. 
In other words, the attenuation process causes the second Gaussian to become wider relative to the first Gaussian in the DGSP (or the first Gaussian becomes narrower in comparison with the second Gaussian), with increasing attenuation. 
Similarly, we observe that there is a tendency for the fluence ratio $f_r$ to decrease with stronger attenuation. 
The energy ratio $w_r^2 f_r$ shows that the 100\% and 58\% cases are similar, while the ratio becomes larger in the 20\% case.
It might indicate that the first Gaussian is more attenuated than the second Gaussian in the 20\% case.

The most pronounced effect that we observe in Table~\ref{tab:att} is the ratio of the transmission and the focal area of the first Gaussian, $T/\Delta^2$:
Decreasing the pulse energy by a factor of five increases this ratio by a factor of five. 
Also note that the ratio changes by almost a factor of two between the two bins of the same data set in the 20\% attenuation case.
Again, it is supposed to be the same ratio in the ideal case where the attenuator would not change the spatial fluence distribution.
Either the transmission through the X-ray focusing optics increases or the focal area of the first Gaussian decreases as the pulse energy decreases, or both change when attenuating the pulse energy.
The reduction of the focal area of the first Gaussian would seem rather surprising because any wavefront distortion induced by the inserted foil would cause the focal spot to become larger, but not smaller.
Thus, we speculate as to the possibility of creating a very irregular focal shape with the hot spot of the beam, which may not be captured by a simple Gaussian profile. 
The increase of the beamline transmission would likely mean that at higher fluences the focusing or transport mirrors are heated up and lose part of their reflectivity.
Even though this effect was found to be small~\cite{Ryutov09}, it may not be completely excluded in our case where high pulse energy (up to 4.5~mJ) was employed.
%
Since $T/\Delta^2$ is inversely proportional to the pulse energy in Eqs.~(\ref{eq:ratio_sg}) and (\ref{eq:ratio_dg}), this conclusion requires the assumption that the experimental attenuation coefficients are correct.
Our calibration cannot determine $T$ and $\Delta$ independently of each other, and both of them are not always well known for different experimental configurations, which currently prevents a detailed understanding of the attenuation behavior.
Nevertheless, our results indicate that the attenuation process does not only reduce the pulse energy, but also it might influence the spatial fluence profile.
In view of these observations, we consider it advisable to calibrate the spatial fluence profile for each and every experimental condition when using attenuated beams.


\section{Conclusion}
\label{sec:conc}
In this work, we have developed the \textsc{xcalib} toolkit 
to calibrate the spatial fluence distribution of an XFEL pulse 
at its focal spot using the CSDs of light atoms. 
The calibration of the spatial fluence profile is essential 
to calculate volume-integrated CSDs and to make a 
quantitative comparison between theoretical 
and experimental results. 
We formulated the calibration procedure based on reinforcement learning
\cite{raschka2015} by using optimization modules in \textsc{python}. 
The automated calibration procedure 
in \textsc{xcalib} is more
efficient than the procedure of manually exploring the parameter space
of the spatial fluence profile performed in previous studies 
\cite{rudek2012,fukuzawa2013,murphy2014,rudek2018}.

Using the \textsc{xcalib} toolkit, we revisited 
the Ar calibrations performed for previous experiments 
\cite{fukuzawa2013,murphy2014,rudek2018}. 
In the first demonstration,
we revisited the Ar calibration at $5.5$ keV employing the SGSP, 
which was performed for a Xe experiment at SACLA 
\cite{fukuzawa2013}. Our result could reproduce the Ar CSD shown in \cite{fukuzawa2013}.
In that reference, the peak fluence was calibrated manually by comparing the yields of ions created by two-photon absorption to the yields of ions created by one-photon absorption. 
In contrast to this manual procedure, we employed 
the cost function defined by the logarithmic difference between
the volume integrated theoretical result and the experimental data.
Using \textsc{xcalib}, it is possible to perform
calibrations without introducing a measure involving such  
physical considerations. 
We also confirmed that the numerical FDF 
of the SGSP agrees with the analytical formula.
In the second demonstration,
we revisited the Ar calibration at $805$ eV performed for
the C$_{60}$ experiment in \cite{murphy2014}. We confirmed that
the low fluence tail modeled by the second Gaussian profile in 
the DGSP is necessary to reproduce the
experimental Ar CSD as demonstrated in \cite{murphy2014}. 
In the third demonstration, we performed the Ar calibration at $6.5$ keV for a recent Xe experiment \cite{rudek2018} to study the 
effect of an attenuator on the functional form of the fluence spatial profile. 
We found that the attenuation process appears to cause a significant modification of the spatial fluence profiles. 
Therefore, when using attenuated beams, a beam profile calibration is advisable for each and every experimental condition.

Our development is essential to automatize the optimization procedure with
flexibility, combining different optimization algorithms, fluence profiles,
charge states, and a wide range of parameter space, which is far
beyond manual procedures employed in \cite{rudek2012,fukuzawa2013,rudek2018}.
Moreover, \textsc{xcalib} has the capability of handling massive amount of
experimental data through automatized optimization procedure.
The calibrated Ar CSDs in this work agree well with
the experimental data except for several charge states  
as previously found in \cite{rudek2012,fukuzawa2013,rudek2018}. 
One way to improve the calibration is improving the 
level of electronic-structure theory being employed 
by including shake-off process and/or double Auger decay 
when computing the ionization dynamics with \textsc{xatom}. 
In the future, \textsc{xcalib} will be employed to a more complex
problem such as calibrating pump and probe pulses. 
\textsc{xcalib} offers us a tool to calibrate such cases with high
efficiency powered by automation. 

\begin{acknowledgments}
H.K.\ and K.U.\ acknowledge that the experiment at SACLA was supported by the X-ray Free Electron Laser Utilization Research Project and the X-ray Free Electron Laser Priority Strategy Program of the Ministry of Education, Culture, Sports, Science and Technology of Japan (MEXT), by the Proposal Program of SACLA Experimental Instruments of RIKEN, by the Japan Society for the Promotion of Science (JSPS) KAKENHI Grant Number JP15K17487, by ``Dynamic Alliance for Open Innovation Bridging Human, Environment and Materials'' from the MEXT, and by the IMRAM project.
N.B., D.R., and A.R.\ acknowledge support from the Chemical Sciences, Geosciences, and Biosciences Division, Office of Basic Energy Sciences, Office of Science, U.S.\ Department of Energy, Grant No.\ DE-SC0012376 (N.B.) and DE-FG02-86ER13491 (D.R.\ and A.R.).
\end{acknowledgments}

\appendix
\section{Flipping the first and second Gaussian profiles}
\label{sec:convert_sol}
Let ${\bf P}=(F_0,f_r,w_r)$ be a set of parameters for the DGSP, 
Eq.~(\ref{eq:dgsp3d}), such that the second Gaussian profile 
is narrower and higher than the first Gaussian, namely, 
$f_r>1$ and $w_r<1$. In such a case, $\tilde{\bf P}=(f_rF_0, 1/f_r, 1/w_r)$ 
gives an equivalent solution having the same cost function value.
The inverses of $f_r$ and $w_r$ make the first Gaussian 
narrower and higher than the second Gaussian profile. The scaling of $F_0 \to f_rF_0$ 
keeps the global peak fluence $F_G$, Eq.~(\ref{eq:FG}), fixed. 
From Eq.~(\ref{eq:ntot_dg}), the global peak fluence $F_G$ 
for the parameter set $\tilde{\bf P}$ is given by
\begin{equation}
F_G(\tilde{\bf P})=F_0+f_rF_0
=\frac{a}{\Delta^2}\frac{1+f_r}{1+w_r^2f_r}\frac{TE}{\omega} \times w_r^2.
\end{equation}
Hence, we also scale the transmission $T$ as $T \to T/w_r^2$, so that $F_G(\tilde{\bf P})$
can satisfy the relation Eq.~(\ref{eq:ntot_dg}) for the parameter set ${\bf P}$.
For the solution $\tilde{\bf P}$, the absolute
ion yield of a charge state $q$, Eq.~(\ref{eq:absyield}), is given by
\begin{eqnarray}
Y^{(+q)}_{\rm theo}(\tilde{{\bf P}})&=&\int \mathcal{Y}^{(+q)}_{\rm theo}(F({\bf r};\tilde{{\bf P}}) )d^3 r \nonumber \\
&=& \int \mathcal{Y}^{(+q)}_{\rm theo}(F({\bf r^\prime};\tilde{{\bf P}}) )d^3 r^\prime \nonumber \\
&=& {\bar Y}^{(+q)}_{\rm theo}({\bf P}).
\label{eq:scaled_yield}
\end{eqnarray}
On the second line, the integration variables are changed 
to ${\bf r^\prime}=(x^\prime,y^\prime,z^\prime)=(w_rx, w_ry, z/w_r^2)$.
After changing the integration variables, 
the functional form of the fluence spatial profile
coincides with that for the parameter set ${\bf P}$.

\section{Grid scheme for the double Gaussian fluence distribution}
\label{sec:grid_scheme}
During optimizations for the DGSP, the width of one Gaussian
could become narrower or wider than
the other, in contrast to the initial guess. Sufficiently fine grid points
must be used to describe the narrower Gaussian to accurately calculate
volume integrations. To handle such unpredictable situations, the grid points
have to be dynamically changed at each optimization steps. The first
step to define our grid points is to find out which Gaussian width is 
narrower or wider, namely, 
\begin{subequations}
\begin{eqnarray}
\Delta_{<}&=&{\rm min}\left(\Delta_1, \Delta_2\right),\\
\Delta_{>}&=&{\rm max}\left(\Delta_1, \Delta_2\right).
\end{eqnarray}
Using the narrower width, we define the grid spacing
for the narrower Gaussian in the interval of 
$\left[-L_{<}/2, L_{<}/2\right]$,
\begin{equation}
l_{<}=\frac{L_{<}}{n_{\rm grid}},
\end{equation}
where $n_{\rm grid}$ is the default number of grid points.
The length of the interval $L_{<}$ is determined 
as small as possible that yet the amplitudes of the narrower 
Gaussian at the borders can satisfy the condition,
\begin{equation}
e^{-\frac{L_<^2}{2\Delta^2_<}} \ll 1.
\label{eq:gaussian_border}
\end{equation}
Using the quantity $l_<$, we define the grid points for
both of two Gaussian profiles,
\begin{equation}
n=\left[\frac{L_>}{l_<}\right],
\end{equation}
\end{subequations}
where the bracket is the operator which returns the nearest integer of
the argument.
In this work, we use $L_{\lessgtr}=7\Delta_{\lessgtr}$, then the left hand side
Eq.~(\ref{eq:gaussian_border}) is $\sim 2.2 \times 10^{-3}$.

\section{Slope value in the low fluence limit}
\label{sec:slope}
Using $\ln(1+x) \approx x$ for $|x| \ll 1$, the slope of the logarithm of
the absolute ion yield at fluence value $F_0$ may be computed via
\begin{equation}
 \ln Y_{{\rm theo}}^{(+q)}(F_0+\Delta F_0)-\ln Y_{{\rm theo}}^{(+q)}(F_0) 
\approx \frac{1}{Y_{{\rm theo}}^{(+q)}(F_0)}\frac{\partial Y_{{\rm theo}}^{(+q)}(F_0)}{\partial F_0}\Delta F_0.
\label{eq:numerator}
\end{equation}
Here the dependency of the absolute ion yield on other parameters is
omitted for simplicity. In the same way,
\begin{equation}
\ln(F_0+\Delta F_0)-\ln F_0 \approx \frac{\Delta F_0}{F_0}.
\label{eq:denominator}
\end{equation}
Using Eqs.~(\ref{eq:numerator}) and (\ref{eq:denominator}), 
the slope, Eq.~(\ref{eq:slope}), is given by
\begin{equation}
s(F_0)=\lim_{\Delta F_0 \to 0} \frac{ \ln Y_{{\rm theo}}^{(+q)}(F_0+\Delta F_0)
-\ln Y_{{\rm theo}}^{(+q)}(F_0)}{\ln(F_0+\Delta F_0)-\ln F_0}
=F_0\frac{\partial}{\partial F_0} \ln Y_{{\rm theo}}^{(+q)}(F_0).
\end{equation}
Substituting the definition of the absolute yield, Eq.~(\ref{eq:absyield}),
\begin{equation}
s(F_0)=F_0\left( \int \frac{\partial \mathcal{Y}_{{\rm theo}}^{(+q)}(F_0)}{\partial F}\frac{\partial F}{\partial F_0} d^3 r\right) 
/ \int \mathcal{Y}_{{\rm theo}}^{(+q)}(F_0) d^3 r. 
\label{eq:slope3}
\end{equation}
We assume that the spatial fluence profile may be written as
\begin{equation}
F({\bf r};{\bf P})=F_0f({\bf r};{\bf p}),
\label{eq:sep}
\end{equation}
where the symbol ${\bf p}$ represents other parameters, namely
\begin{equation}
{\bf P}=(F_0, {\bf p}).
\end{equation}
In the low fluence limit, substituting 
\begin{equation}
\mathcal{Y}^{(+q)}_{\rm theo}(F({\bf r};{\bf P})) \propto [F({\bf r};{\bf P})]^{n_q}
\label{eq:LOPT}
\end{equation}
and Eq.~(\ref{eq:sep}) into Eq.~(\ref{eq:ionyield}), it may be shown that
\begin{equation}
\lim_{F_0 \to 0}s(F_0)=n_q.
\label{eq:slope2}
\end{equation}

\section{Fluence distribution function}
\label{sec:flu_dist_deriv}
The definition of fluence distribution function is given by
\begin{equation}
V(f)=\int \delta\left(F({\bf r})-f\right)d^3 r. 
\end{equation}
Switching to polar coordinates, and assuming $F({\bf r})=F(r)$,
\begin{equation}
V(f)=2\pi\int_0^\infty \frac{\delta(r-r_f)}{|F^\prime(r_f)|}rdr,
\end{equation}
where $r_f$ satisfies $F(r_f)=f$.
For the SGSP,
\begin{equation}
F^\prime(r_f)=-\frac{2\pi a r_f}{\Delta^2}f. 
\end{equation}
We thus obtain
\begin{equation}
\label{eq:flu_dist_sg}
V(f)=
\left\{
\begin{array}{cc}
\frac{\Delta^2}{a}\frac{1}{f} & (~f\leq F_0~), \\
0 & (~f>F_0~).
\end{array}
\right.
\end{equation}
Note that Eq.~(\ref{eq:flu_dist_sg}) depends on the fit parameter $F_0$
only through cutoff condition at $F_0$.

In the case of the DGSP, 
there is no analytical formula for the fluence distribution function. 
However, it is still possible to understand its overall 
behavior considering the model,
\begin{equation}
f_r \ll 1~~~{\rm and}~~~~w_r \gg 1.
\label{eq:asymptotic_condition}
\end{equation}
This model was employed before \cite{murphy2014} to describe  
the halo in the spatial profile of XFELs utilizing
the second Gaussian profile.
First we consider the solution $r_f$ for a fluence value $f$ 
much smaller than the peak fluence of the second Gaussian profile $f_rf$,
such that $f \ll f_rF_0$. In such a situation the first Gaussian can 
be ignored. Then the solution $r_f$ roughly satisfies
\begin{eqnarray}
F(r_f) &\approx& f_rF_0e^{-\pi a \frac{r_f^2}{(w_r\Delta)^2}}, \\
F^{\prime}(r_f)&\approx&-\frac{2\pi a}{(w_r \Delta)^2}r_ff.
\end{eqnarray}
Therefore, following the derivation for the SGSP,
we obtain 
\begin{equation}
V(f) \approx \frac{(w_r\Delta)^2}{a}\frac{1}{f}~(~f \ll f_rF_0~) .
\label{eq:occ_density_dg1}
\end{equation}
It is important to realize that the result Eq.~(\ref{eq:occ_density_dg1})
depends on the additional fit parameters $w_r$. Therefore comparing with
the single Gaussian case, in that sense, the behavior of the 
fluence distribution function for small
fluence values is not universal. 
The fluence distribution function for a fluence value $f \gg f_rF_0$
can be also derived ignoring the second Gaussian profile.
The result coincides with that of Eq.~(\ref{eq:flu_dist_sg}) 
in the single Gaussian case. 

\bibliography{xcalib}

\end{document}